\documentclass{ceab}   

\usepackage{epsfig}     
\usepackage{graphicx}   
\usepackage{ceabbib}     
\usepackage[T1]{fontenc}

\newcommand{\ep}{$\varepsilon$~Aur }
\newcommand{\epe}{$\varepsilon$~Aur}

\newcommand{\fotel}{{\tt FOTEL} }

\newcommand{\korel}{{\tt KOREL} }

\newcommand{\spefo}{{\tt SPEFO} }

\newcommand{\ubv}{\hbox{$U\!B{}V$}}
\newcommand{\bvri}{\hbox{$B{}V\!R{}I$}}
\newcommand{\bv}{\hbox{$B\!-\!V$}}
\newcommand{\ub}{\hbox{$U\!-\!B$}}

\newcommand{\oc}{\hbox{$O\!-\!C$}}

\newcommand{\p}{$\pm$}

\newcommand{\D}{$^{\rm d}\!\!.$}

\newcommand{\kms}{km~s$^{-1}$ }
\newcommand{\ks}{km~s$^{-1}$}

\newcommand{\ms}{M$_{\odot}$}

\newcommand{\ha}{H$\alpha$ }

\newcommand{\hae}{H$\alpha$}

\newcommand{\siii}{Si~II~6347~\AA\ }
\newcommand{\siiie}{Si~II~6347~\AA}
\newcommand{\feii}{Fe~II~6417~\AA\ }
\newcommand{\feiie}{Fe~II~6417~\AA}

\begin{document}
\countdef\pageno=0
\pageno=99

\def\tit{A new look into the spectral and light variations of \epe}
\def\aut{P. HARMANEC et al.}
\def\str{99--108}

\title{A new look into the spectral and light variations of \epe}

\author{P. HARMANEC$^1$, H. BO\v{Z}I\'C$^2$, D.~KOR\v{C}\'AKOV\'A$^1$,
L.~KOTKOV\'A$^3$,\\
P.~\v{S}KODA$^3$, M.~\v{S}LECHTA$^3$, M.~\v{S}VANDA$^{1,3}$,
V.~VOTRUBA$^3$,\\
M.~WOLF$^1$, P.~ZASCHE$^1$, A.~HENDEN$^4$ and J.~RIBEIRO$^5$
\vspace{2mm}\\
\it $^1$Astronomical Institute of the Charles University,\\
\it Faculty of Mathematics and Physics,\\
\it V~Hole\v{s}ovi\v{c}k\'ach~2, CZ-180~00~Praha~8, Czech Republic\\
\it $^2$Hvar Observatory, Faculty of Geodesy, University of Zagreb,\\
\it Ka\v{c}i\'{c}eva 26, HR--10000 Zagreb, Croatia\\
\it $^3$Astronomical Institute, Academy of Sciences of the Czech Republic,\\
\it CZ-251 65 Ond\v rejov, Czech Republic\\
\it $^4$AAVSO, 49 Bay State Road, Cambridge, MA 02138, USA\\
\it $^5$Observat\'orio do Instituto Geogr\'afico do Ex\'ercito,\\
\it R.~Venezuela 29, 3 Esq. 1500-618, Lisboa, Portugal
}

\maketitle

\begin{abstract}
Investigating long series of spectral and photometric observations, we
found that the orbital elements of \ep are subject to much
larger uncertainties than usually believed. The \ha emission is found to
move basically with the F primary but its exact location should still be
investigated. We also find strong additional absorption and large
reddening of the object near the third contact during the eclipse.
Episodic atmospheric mass transfer from the F primary towards its companion
is tentatively suggested.
\end{abstract}


\section{Observations}
Having at our disposal rich series of spectral and photometric observations
of \epe, we attempted to find out what are the safe and uncertain
observational facts about this intriguing binary and provide also some
speculations about its true nature. In particular, we analyze
the following sets of spectra from the red spectral region covering \hae:
112 DAO CCD spectra from 1994 -- 2011; 292 Ond\v{r}ejov spectra from
2006 -- 2012; and 15 Lisbon spectra from 2011 -- 2012.
Systematic \ubv\ photometry, carefully calibrated to the standard Johnson
system was obtained at Hvar and combined with the \bvri\ photometry secured
by AH in the framework of the AAVSO program.

\vspace{-2.5mm}
\section{The timing of the last eclipse}
\vspace{-1mm}
All time instants in this paper are in RJD=HJD-2400000.0 (HJD being
the heliocentric Julian date of observation).
The photometric eclipse lasted from RJD 55050 to 55800;
the epoch of the primary mid-eclipse being RJD~55403
\citep{chadima2010}, the second contact RJD~55225
and the third contact RJD~55620. The spectroscopic eclipse as seen in
the \ha absorption started, however, {\sl three years earlier}
\citep{chadima2011}.

\vspace{-2.5mm}
\section{What can be learned from the dynamical spectra?}
\begin{figure}
\begin{minipage}[b]{0.45\linewidth}
\centering
\includegraphics[width=\textwidth]{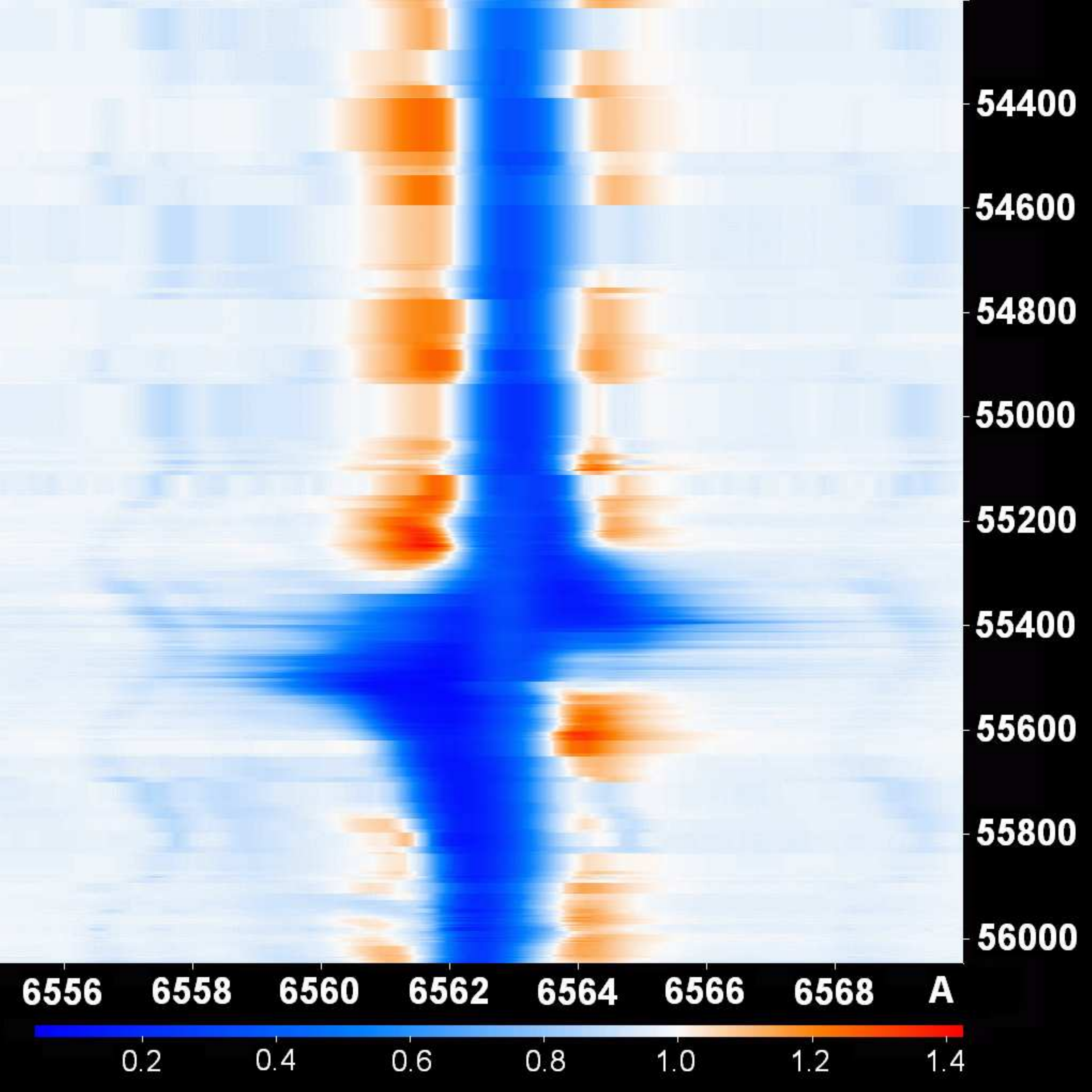}
\caption{The dynamical spectra of the \ha line profile over the time
interval prior to, and during the recent eclipse. The sequence of events
seems
to be similar to that observed in \ha by \citet{wright58}.}\label{hadyn}
\end{minipage}
\hspace{0.5cm}
\begin{minipage}[b]{0.45\linewidth}
\centering
\includegraphics[width=54mm]{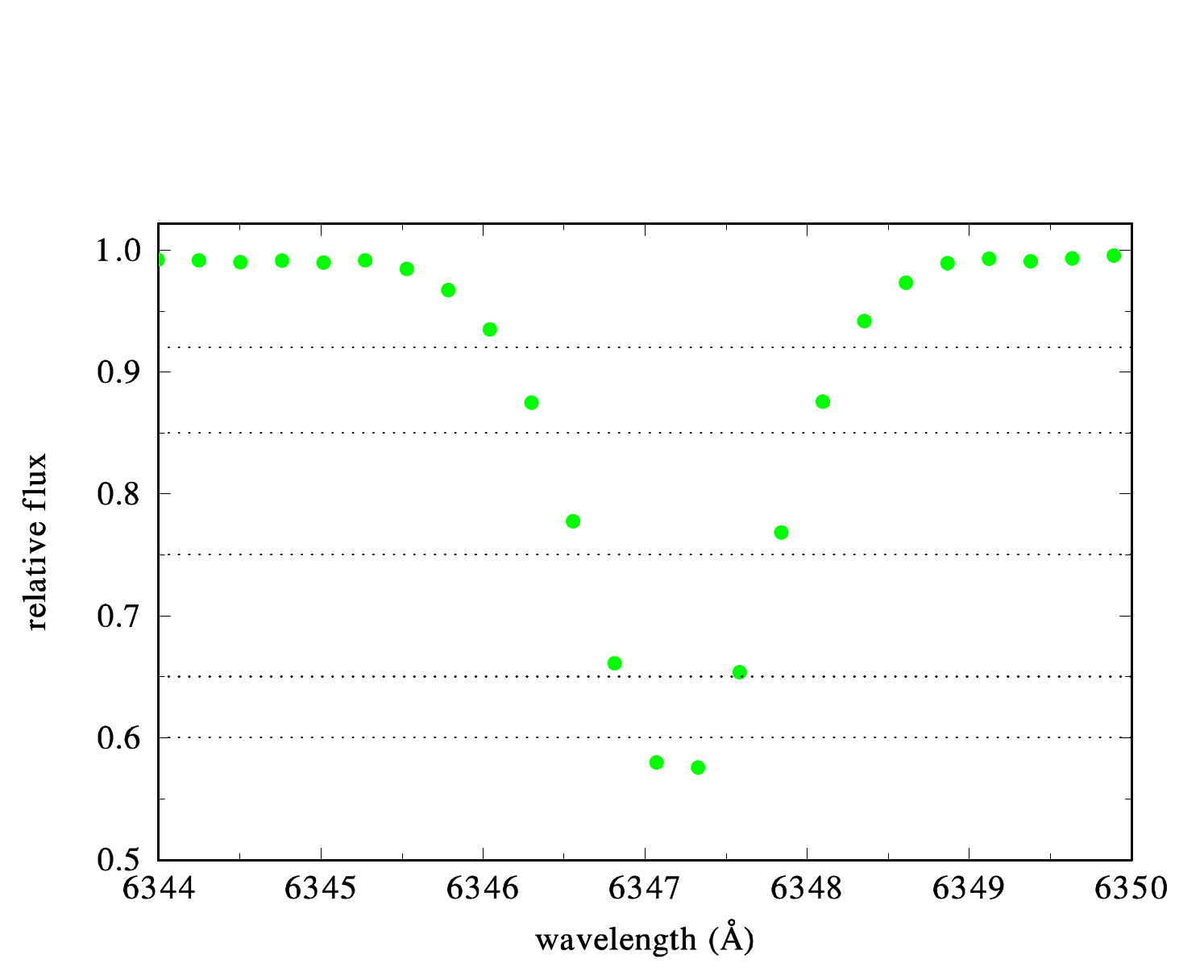}
\caption{A line profile of \siii with relative intensity levels, at which
the local RVs were measured, shown by dotted lines.
}\label{levels}
\end{minipage}
\end{figure}

\vspace{-1mm}
We first investigated the dynamical spectra of three strongest unblended
lines in the red spectral region. Figure~\ref{hadyn} shows the evolution
of the \ha profile. Three facts are worth noting: 1.~The appearance of
additional \ha absorption on the red and then blue side of the profile
{\sl is not symmetric} around the mid-eclipse. The blue-shifted absorption
after the mid-eclipse is stronger and more extended than the red-shifted one
prior to mid-eclipse. 2.~The \ha emission is completely masked by this
extended
absorption. 3.~While the emission exhibits cyclic changes in the $V/R$ ratio
of the violet and red peaks, one can note that prior to eclipse, the $V$
peak
of the emission was statistically stronger than the $R$ peak.

\begin{figure}
\begin{center}
\includegraphics[width=0.45\textwidth]{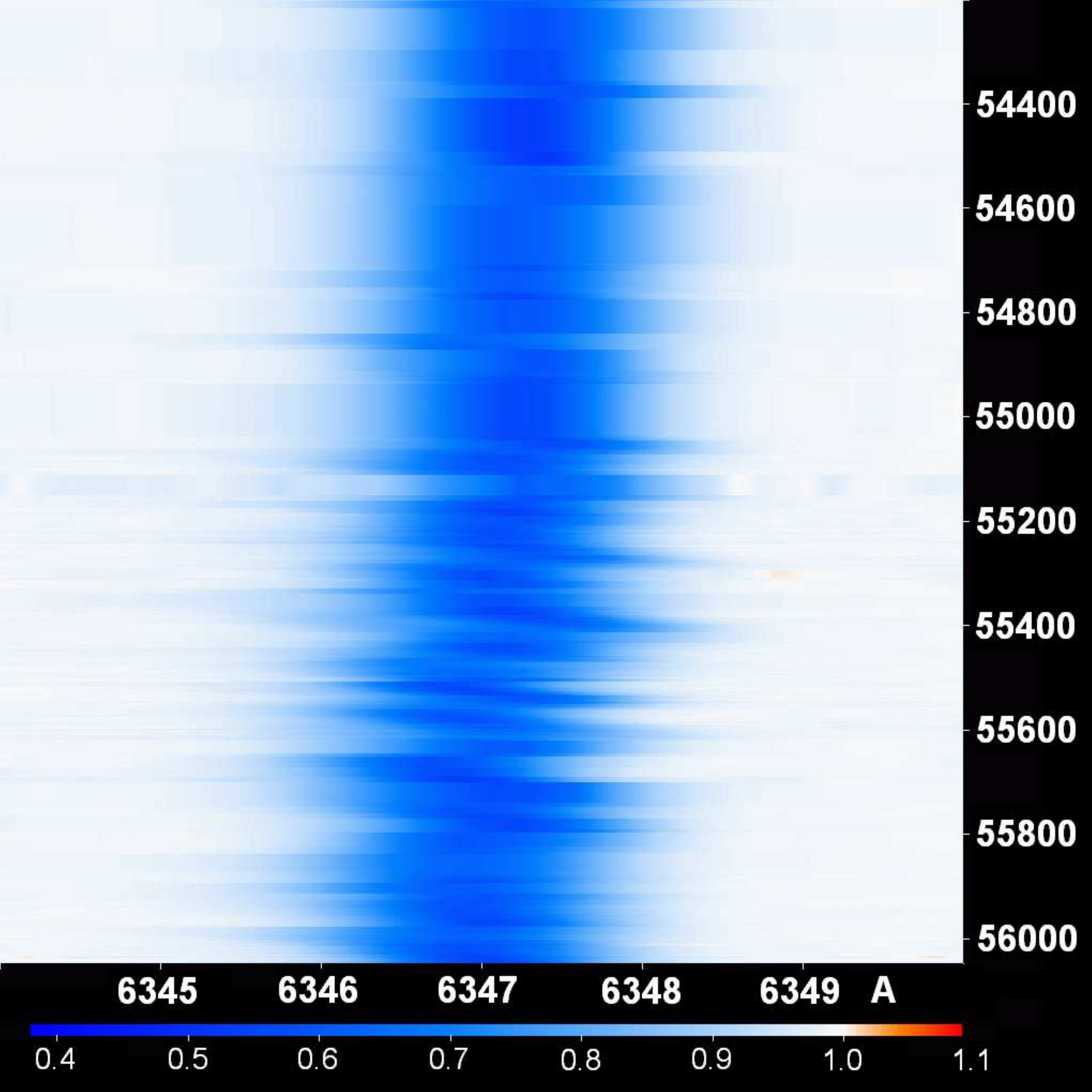}
\includegraphics[width=0.45\textwidth]{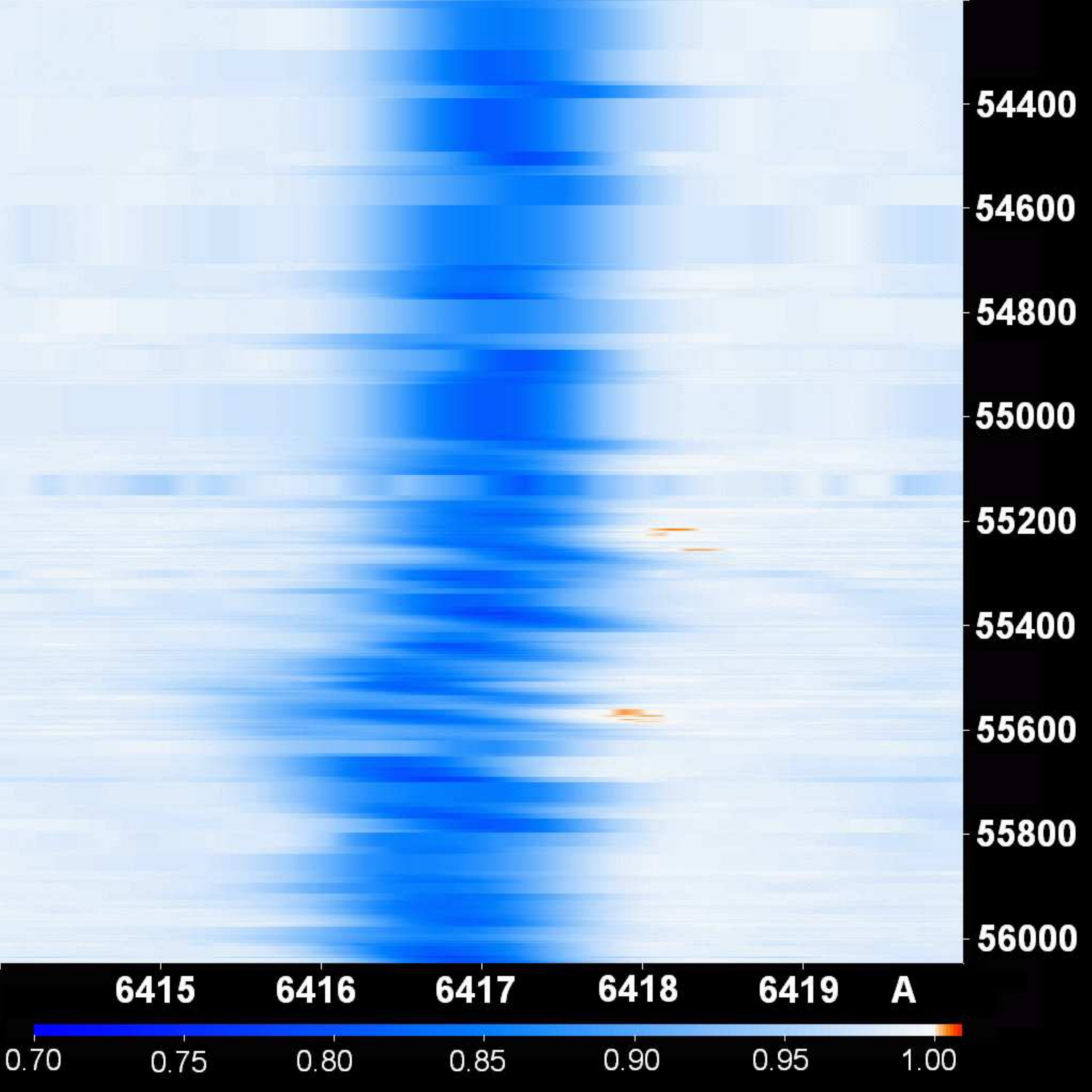}
\caption{The dynamical spectra of the \siii (left) and \feii (right)
line profiles over the time interval prior to, and during the recent
eclipse.}
\label{sifedyn}
\end{center}
\end{figure}

A comparison of the dynamical spectra of two strong metallic lines, \siii
and \feii (cf. Figure~\ref{sifedyn}) reveals two interesting things:
1.~While the general behavior of both lines shows similar patterns, one
can note an extended blue-shifted absorption in phases near the third
contact (RJD~55600) seen in the \feii line only. The inspection of the
spectra shows that there is actually a line doubling seen in \feiie.
2.~In the course of the eclipse, one can see indications of some travelling
subfeatures appearing in the blue wings of both lines and moving across
the line profiles to the red, probably due to rotation of the medium
responsible for the absorptions. A~prototype of such variations
is $\varepsilon$~Per -- \citep[see Figs.~4 to 8 in][]{gies88}.

\vspace{-2.5mm}
\section{Orbital and physical RV changes and
the problem of true orbital elements}
\vspace{-1mm}
To investigate the effect of physical RV variations, which
have an amplitude comparable to the orbital RV changes, local RVs at
various relative intensity levels were measured in the clean profiles of
\siii and \feii
lines (cf. Figure~\ref{levels}). The choice of levels was dictated by the
limited
resolution 11700 of the Ond\v{r}ejov spectra.

\begin{figure}[!ht]
\centering
\includegraphics[width=58.5mm]{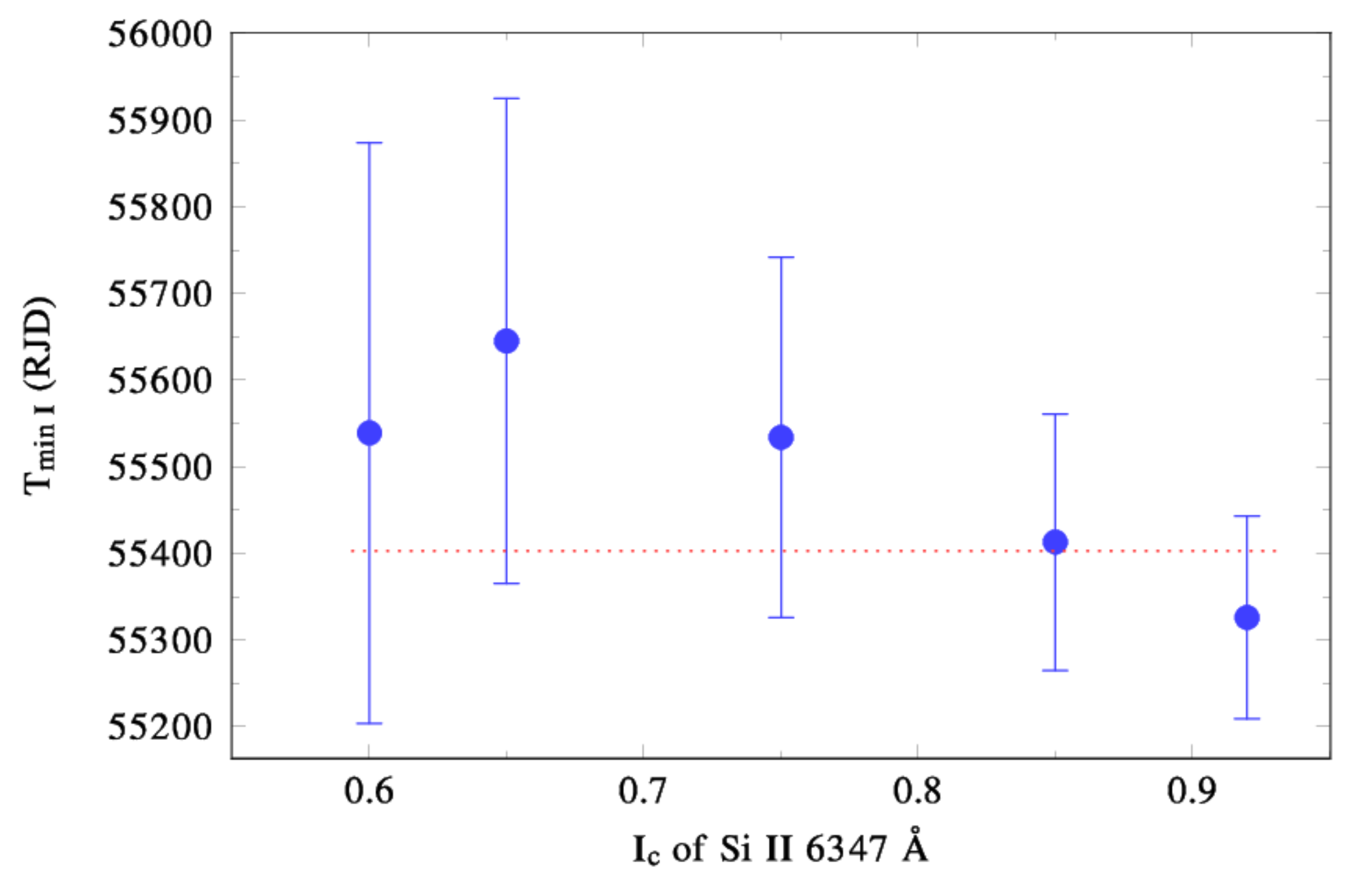}
\includegraphics[width=58.5mm]{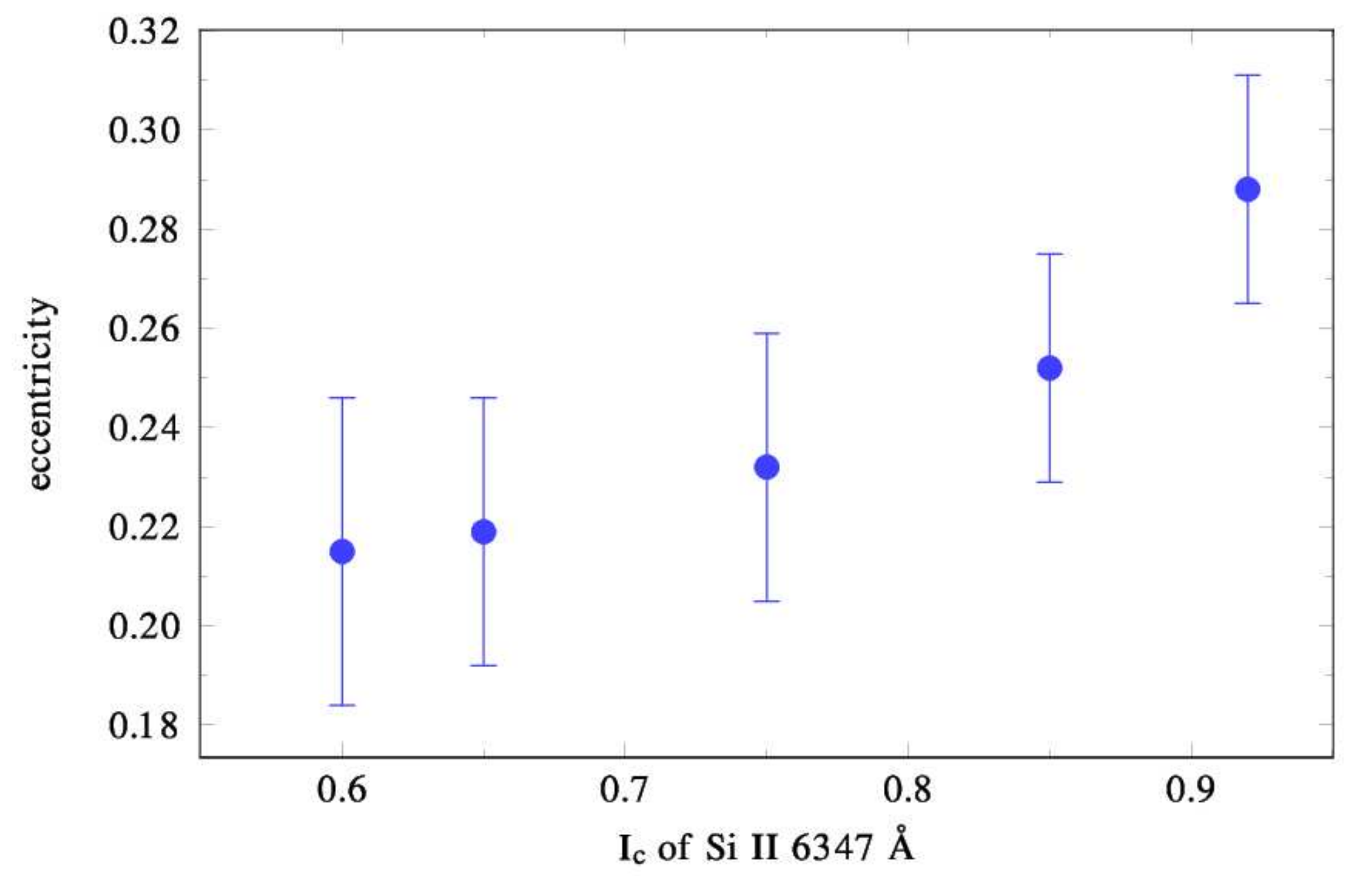}
\includegraphics[width=58.5mm]{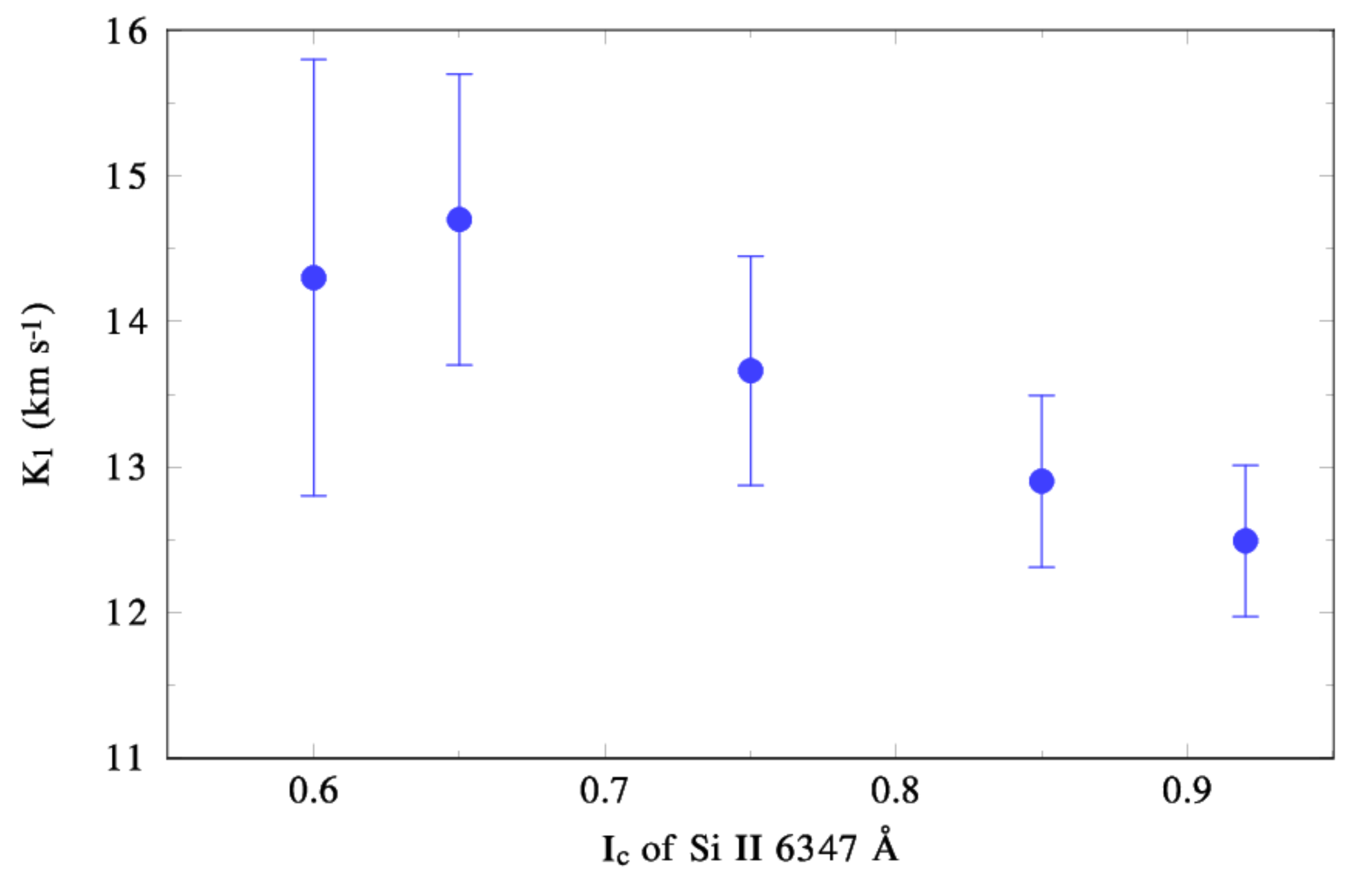}
\includegraphics[width=58.5mm]{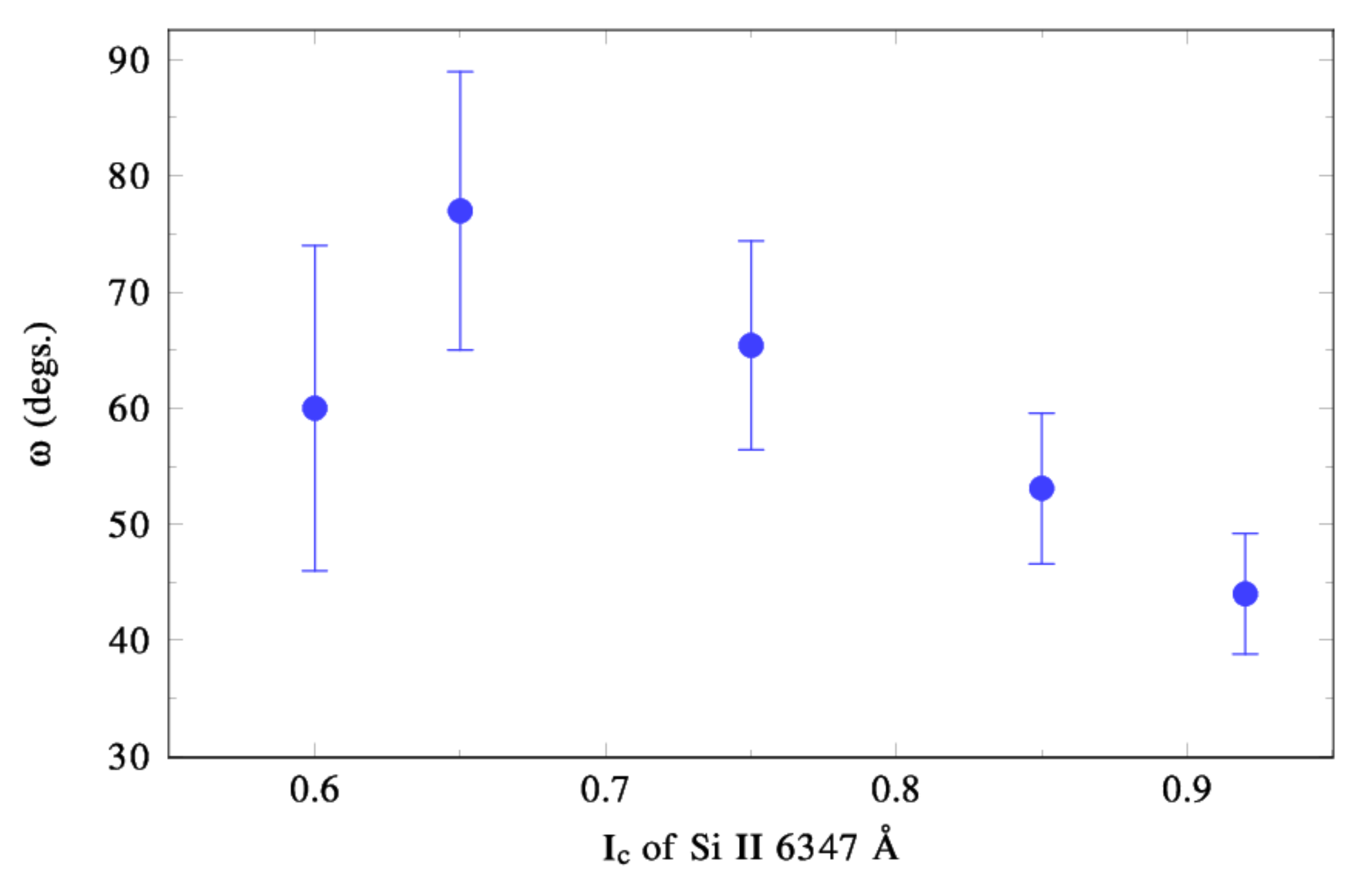}
\includegraphics[width=58.5mm]{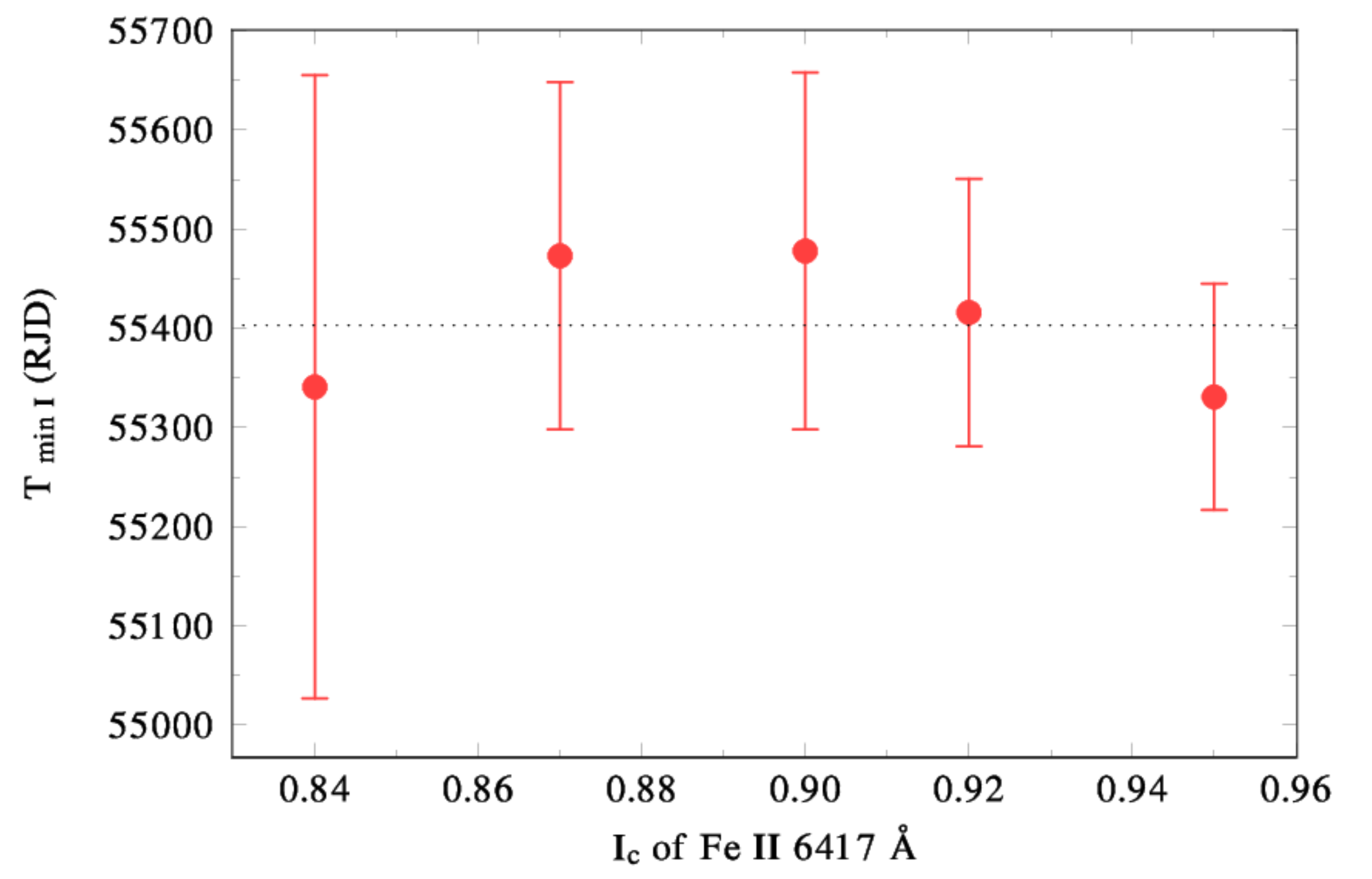}
\includegraphics[width=58.5mm]{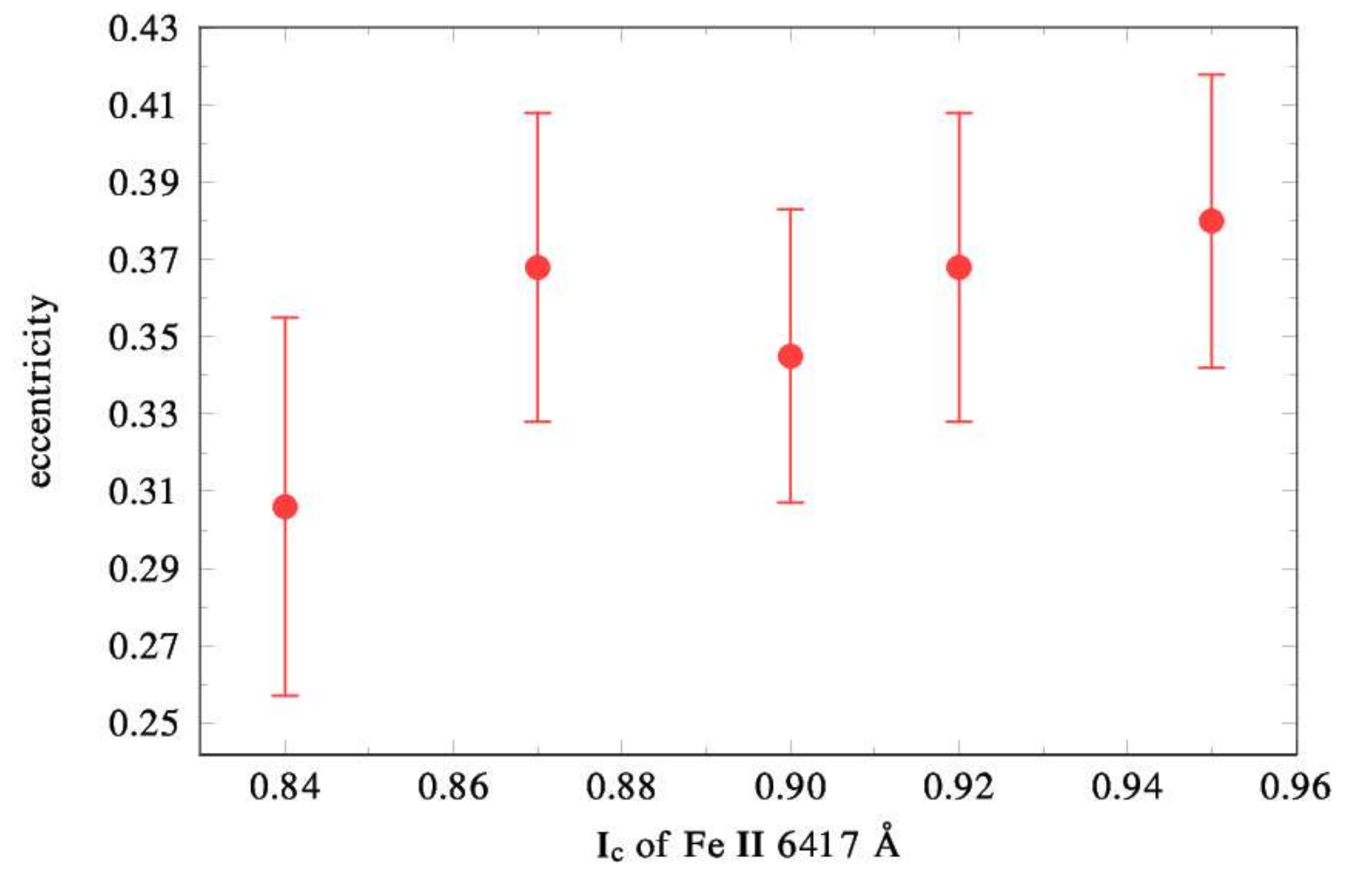}
\includegraphics[width=58.5mm]{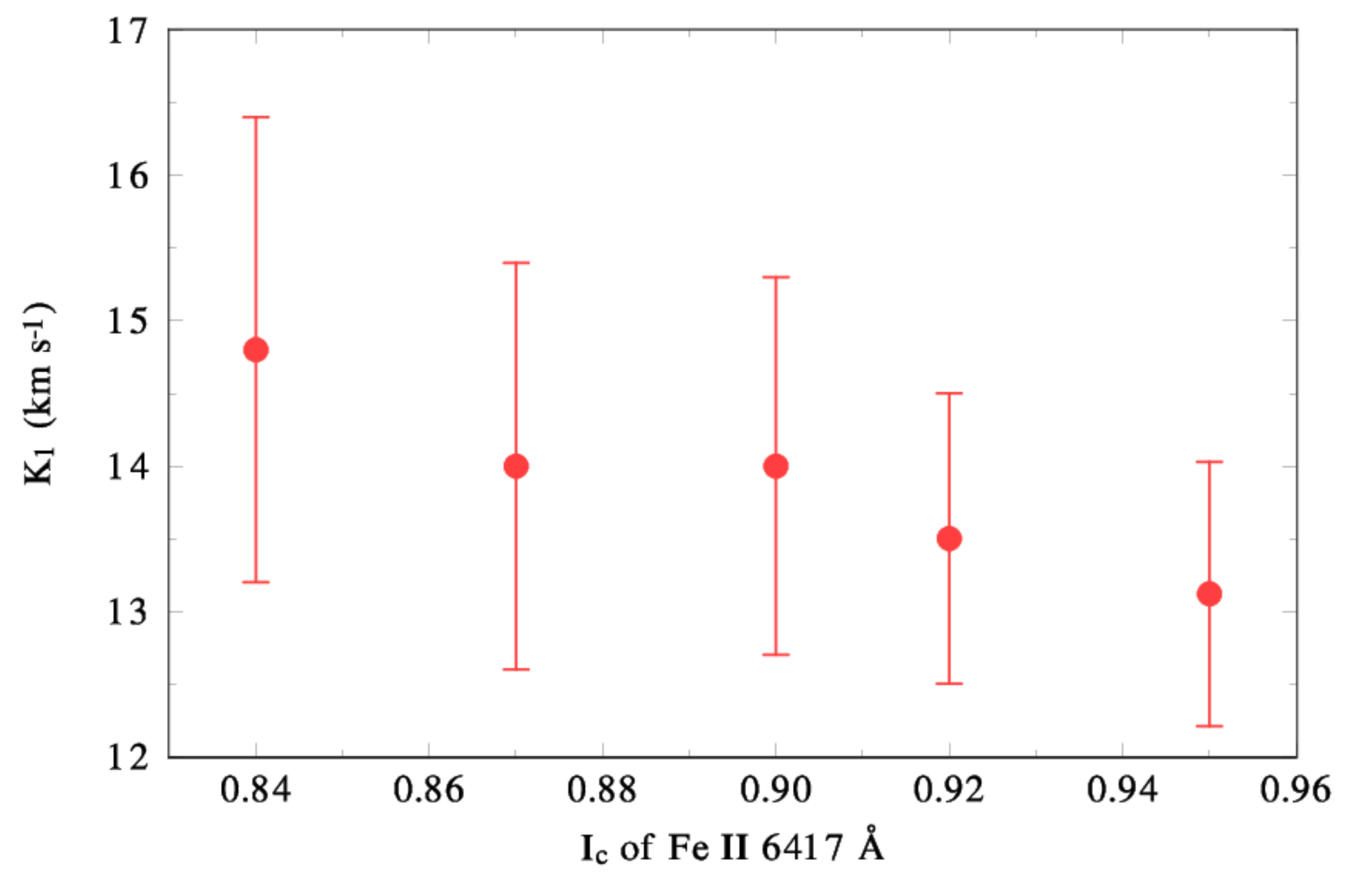}
\includegraphics[width=58.5mm]{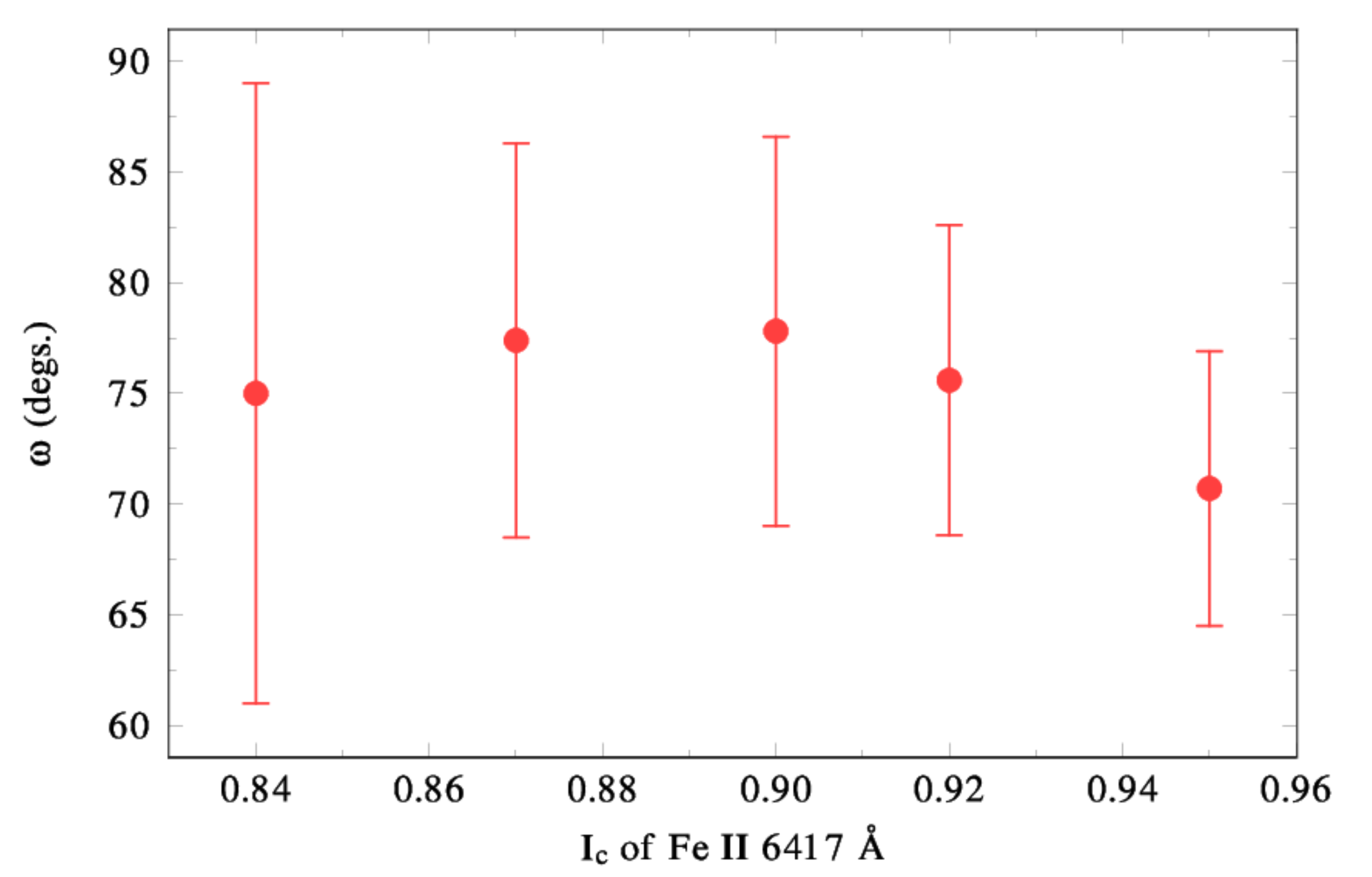}
\vspace{-4mm}
\caption{Evolution of orbital elements with the level, at which the bisector
RV was measured. Top four panels: \siiie, bottom four panels: \feiie.}
\label{local}
\vspace{-5mm}
\end{figure}

First, we derived the bisector RVs at various intensity levels for both
lines,
averaging the local RVs from the blue and red wing measured at the same
level.
Then, the orbital solutions were derived with the program \fotel
\citep{fotel2} for the bisector RVs measured at various levels of the
profiles.
The results, shown graphically in Figure~\ref{local} indicate that all orbital
elements may vary systematically (though the associated errors are not
negligible) over a wide range depending
on the depth within the line profiles where the bisector RV was measured!
In particular, the predicted epoch of the mid-eclipse agrees well with the
observed one (RJD~55400) only for RVs derived on the outer wings
of the profiles. Also the semiamplitude, eccentricity and orientation
of the orbit vary quite substantially. This may have important consequences.
First, since the orientation of the orbit is {\sl poorly constrained}, one
should look for the secondary minimum in the far IR systematically
and without prejudice, it might come already some 15 years after
the primary mid-eclipse! Its firm detection would help to constrain the true
orientation of the orbit. Note also that the difference between
$K_1=9.1$~\kms (Table~I) and 14.8~\kms (Figure~4) is huge for the long orbital
period of \epe: For instance, for a primary mass of 20~\ms, the
corresponding
range of mass ratios following from the mass function is 0.42 to 0.77,
implying the secondary mass between 8.4 and 15.3~\ms.

\begin{table}
\vspace{-3mm}
\caption{Orbital elements of \ep based on the RVs of the wings of \ha
emission
and on the RVs of weak symmetric absorption lines. The orbital period was
kept
fixed at 9890\D26 \citep{chadima2010}.}
\label{elem}
\vskip 2mm
\begin{tabular}{lccc}
\hline
Element& \ha emission wings & weak symmetric absorptions\\
\hline
$T_{\rm min.I}$ (RJD)      &55526\p383&55389\p173\\
$T_{\rm min.II}$ (RJD)     &   61485  & 61194\\
$e$                        &0.167\p0.033&0.301\p0.038\\
$\omega$ (deg.)            &15\p15   & 64.1\p8.1\\
$K_1$ (\ks)                &9.1\p1.2 & 13.28\p0.95\\
$\gamma$ (\ks)             &$-$1.60\p0.34&$-$2.26\p0.56\\
rms 1 obs. (\ks)           &3.55&5.67\\
\hline
\end{tabular}
\end{table}

\begin{figure}
\centering
\vspace{-3mm}
\includegraphics[width=0.49\textwidth]{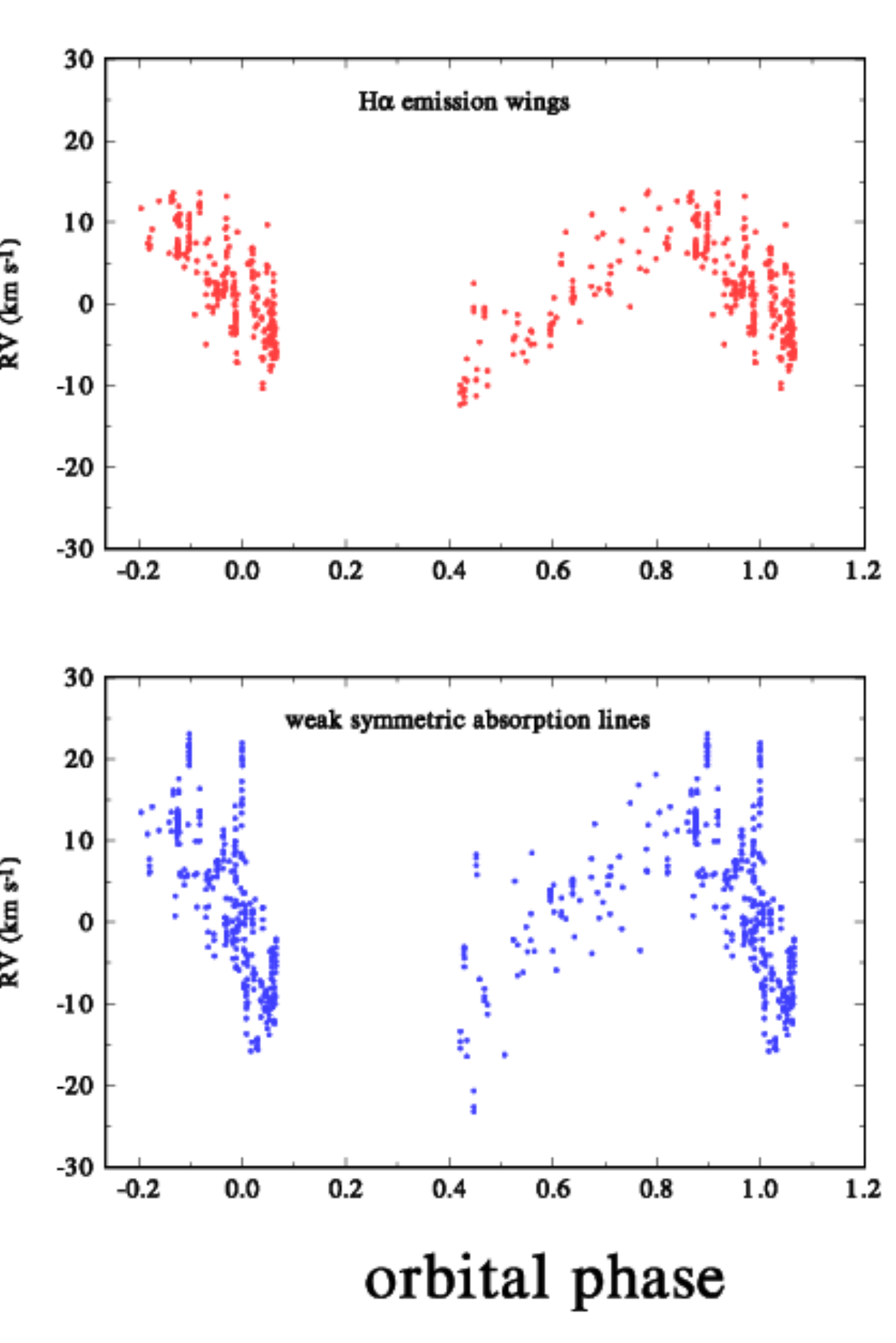}
\includegraphics[width=0.49\textwidth]{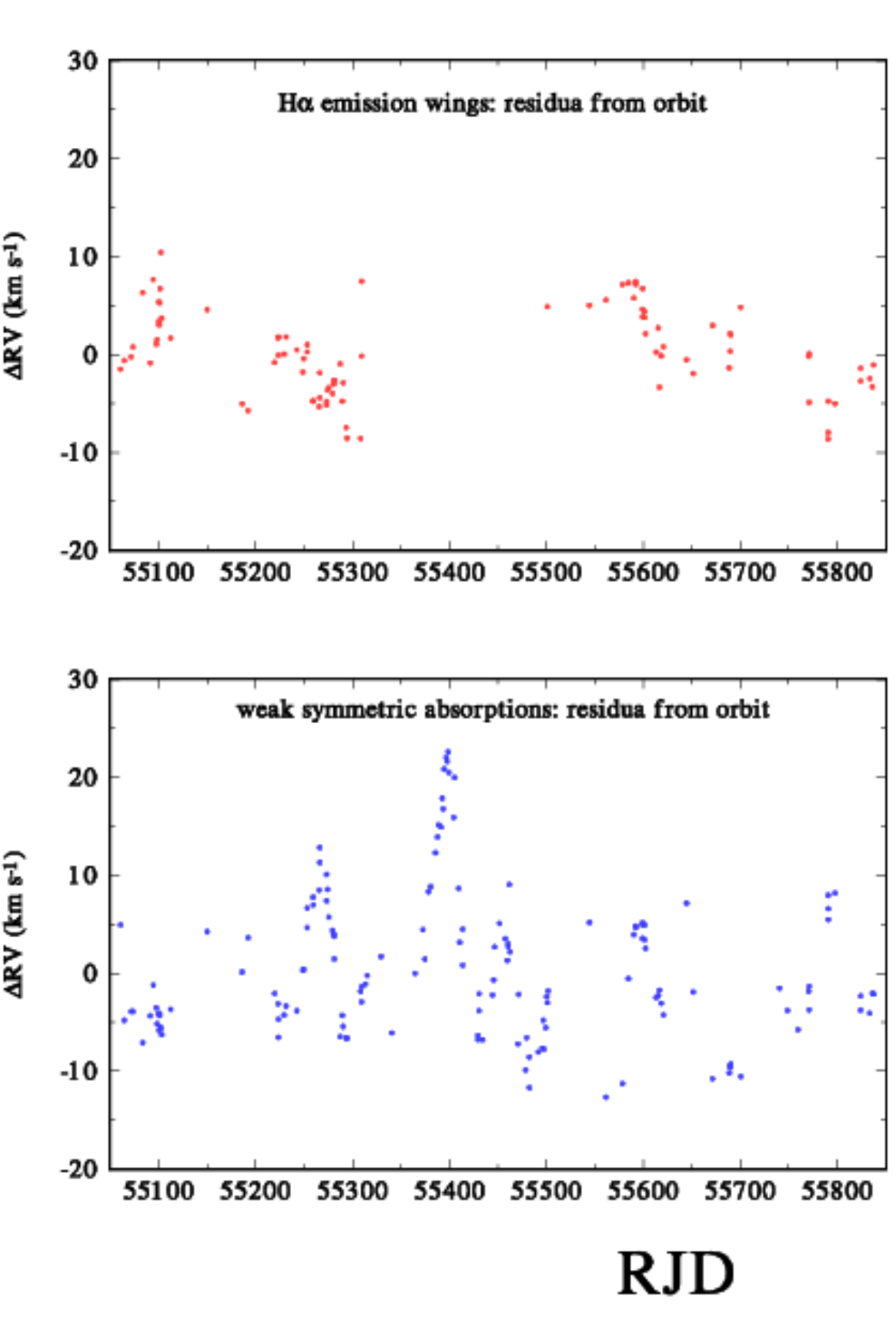}
\vspace{-6mm}
\caption{{\sl Left panels:} RV curves based on the wings of the \ha emission
and on symmetric weak absorption lines.  {\sl Right  panels:}
The O-C deviations from the orbital solutions. See the text for details.}
\label{rvcs}
\vspace{-4mm}
\end{figure}

We also derived RV curve of the \ha emission wings and of weak
symmetric absorption lines, which we noted in the spectra.
To this end, we first derived a mean out-of-eclipse
spectrum disentangled with the program \korel \citep{korel3}. Using the
program \spefo \citep{sef0,spefo}, we compared the disentangled line
profiles with the individual observed spectra on the computer screen
and recorded the needed RV shifts to bring the profiles to overlap.
In other words, we carried out something like a `manual cross-correlation'.
The advantage of this tedious procedure is that one can avoid the telluric
lines and flaws. Our result constitutes the final proof that the \ha
emission
moves basically with the F primary. The orbital elements based on
the wings of the \ha emission and on the mean RV of weak symmetric
absorptions
are compared in Tab.~\ref{elem} and the corresponding orbital RV curves
and the \oc\ residua from them are shown in Figure~\ref{rvcs}.
Note the difference in the amplitude and shape of the curves as well as in
the pattern of residual physical RV changes.

\vspace{-2mm}
\section {Light and colours changes}
\begin{figure}
\vspace{-2mm}
\begin{minipage}[t]{0.49\linewidth}
\includegraphics[width=90mm,angle=-90]{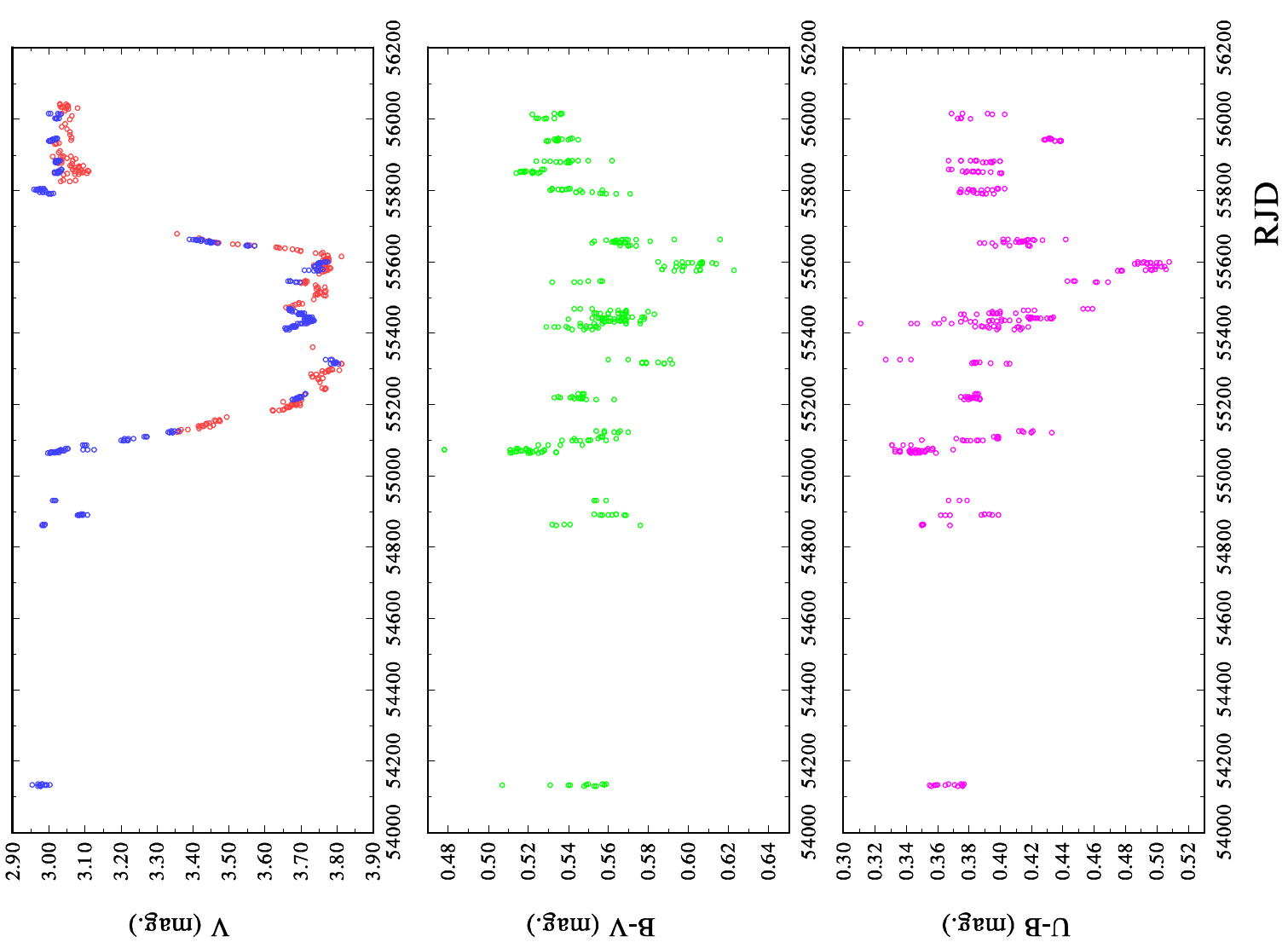}
\caption{The $V$ light curve and \bv\ and \ub\ colour
changes during the last eclipse.}
\label{ubv}
\end{minipage}
\hspace{0.5cm}
\begin{minipage}[t]{0.49\linewidth}
\centering
\includegraphics[width=90mm,angle=-90]{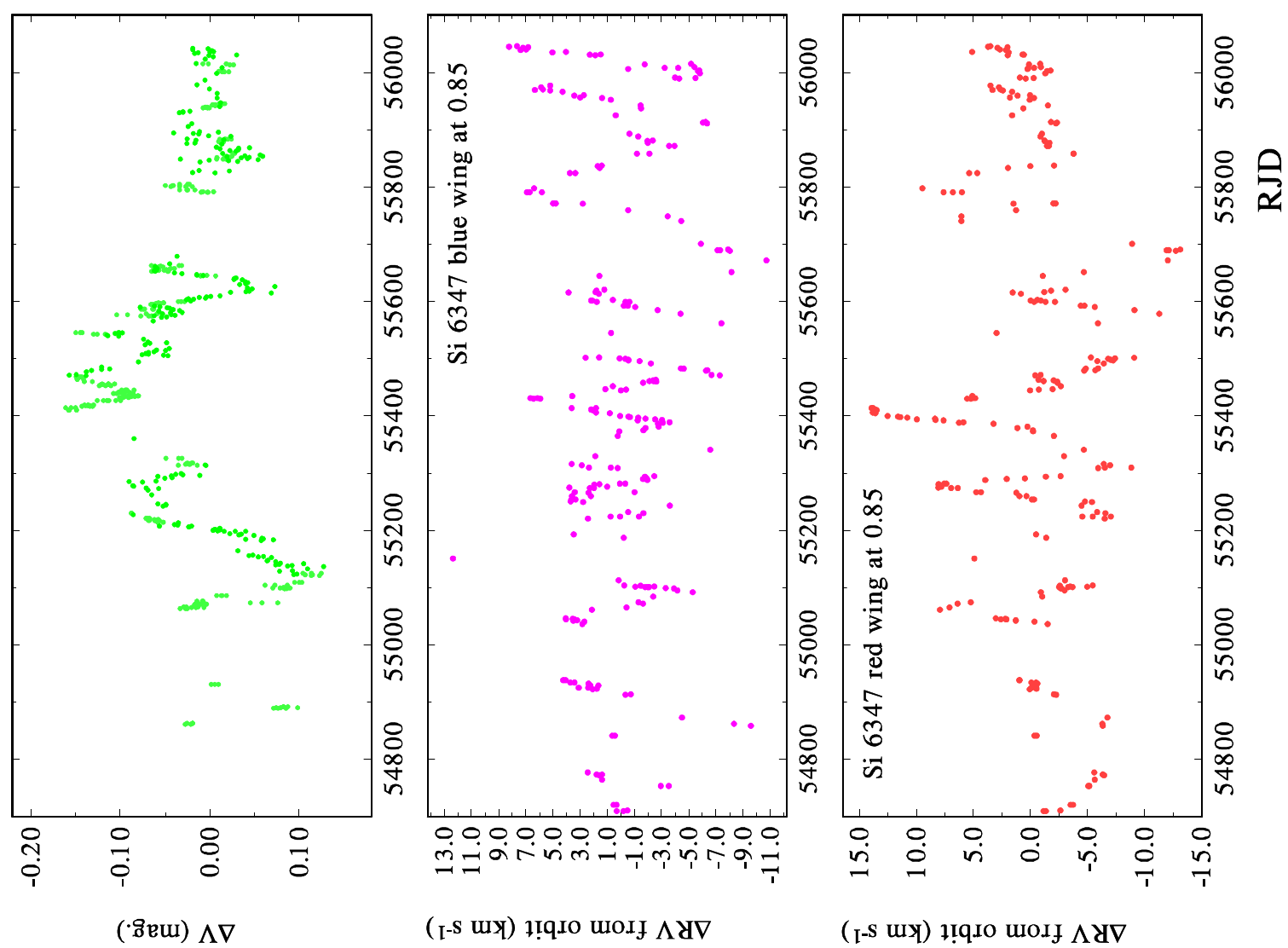}
\vspace{-5mm}
\caption{The \oc\ deviations
from the $V$ light curve compared to the \oc\ deviations from the locally
measured RVs on the left and right wings of the \siii line.}
\label{ocvsi}
\end{minipage}
\vspace{-2mm}
\end{figure}

\vspace{-1mm}
In Figure~\ref{ubv} we reproduce our light and colour curves of \epe.
It is well seen that the physical light variations continued
even during the whole eclipse. A very
interesting new finding is the pronounced reddening of both colour indices
near the third contact (RJD~55600), i.e. at the same time interval
when the extended blue-shifted contribution is seen in the \feii absorption
line (cf. Figure~\ref{sifedyn}). In Figure~\ref{ocvsi} we compare the physical
light variations in $V$ (removing variations due to the eclipse)
with the physical
RV variations of local velocities measured at the blue and red
wing of the \siii line at relative intensity 0.85 (removing
the orbital changes). One can see that the variations at both line wings
are usually in phase with each other but have different amplitudes, their
correlation coefficient being 0.80. There might also be some correlation
of local RVs with the light changes.

\vspace{-2mm}
\section{The dark disk: Ring-like structure or corotating inhomogeneities?}
\begin{figure}
\begin{minipage}[b]{0.45\linewidth}
\centering
\includegraphics[width=55mm]{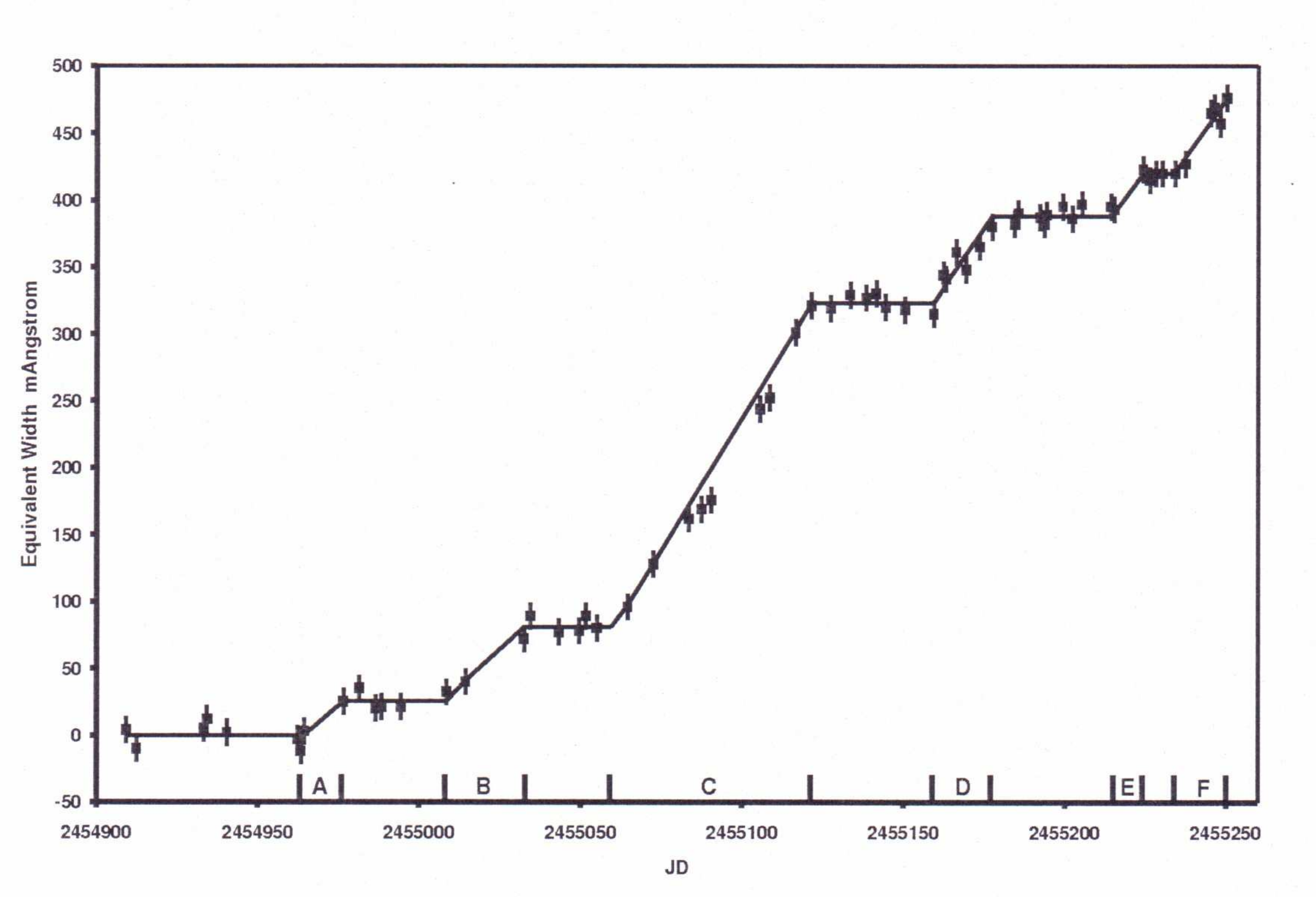}
\caption{A reproduction of the plot of the equivalent width of the
K~I~7699~\AA\
line from the paper by \citet{lead2010}. The authors draw a step function
and
interpret it as evidence of a ring-like structure of the disk around the
secondary.}
\label{lead}
\end{minipage}
\hspace{0.5cm}
\begin{minipage}[b]{0.45\linewidth}
\centering
\includegraphics[width=55mm,angle=-90]{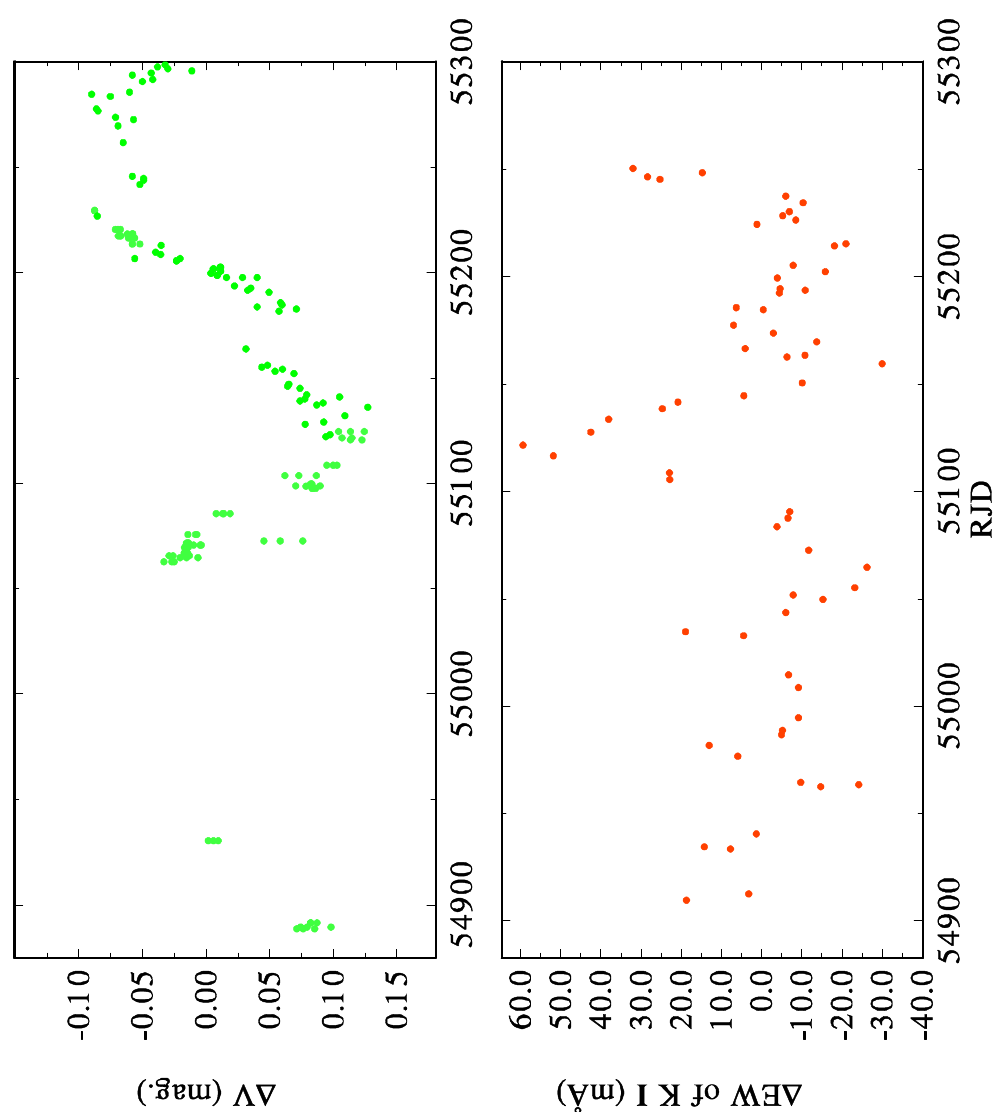}
\caption{The \oc\ from the K~I equivalent width after detrending for a
secular trend compared to the \oc\ deviations from the eclipse light curve.}
\label{leadoc}
\end{minipage}
\end{figure}

\vspace{-1mm}
\citet{lead2010} argued that the excess EW of the K I absorption line
during the eclipse varies in steps (see Figure~\ref{lead}, which is
the reproduction of their Figure~3 with a step function drawn)
and elaborated on the idea by \citet{fer90} that the disk might be actually
a system of concentric rings reminiscent of Saturn's rings.

We removed the trend from their EWs via spline functions and plot
the \oc\ deviations, comparing them with the $V$ photometry residua
(eclipse changes removed) from the same
time interval in Figure~\ref{leadoc}.
Both variations are {\sl smooth (not in steps)}, cyclic and probably
mutually
correlated.  This might support the idea of corotating structures
(spokes? resonantly excited density waves?) in the disk, instead of
the suggested ring structure.

\section{What we learned and how to interpret it?}
\begin{itemize}
\itemsep -2pt
\item The \ha emission moves in orbit with the F primary but
with a smaller RV amplitude than that obtained from the absorption lines.
\item The physical variations during the eclipse seem to be
governed by two characteristic timescales
\citep[the first one already noted in some previous studies,
e.g.][]{kim2008,
chadima2011}: cyclic changes
on a time scale of 66~days and the travelling subfeatures moving from the
blue
to the red edge of the line profiles, which were found by us.
\item Compelling evidence of extra circumstellar material projected
against the primary near phases of the third contact around RJD 55600 is
provided by a large reddening in the \ub\ and \bv\ indices, light decrease,
and doubling of the \feii line.
\end{itemize}

\begin{figure}
\centering
\includegraphics[width=45mm]{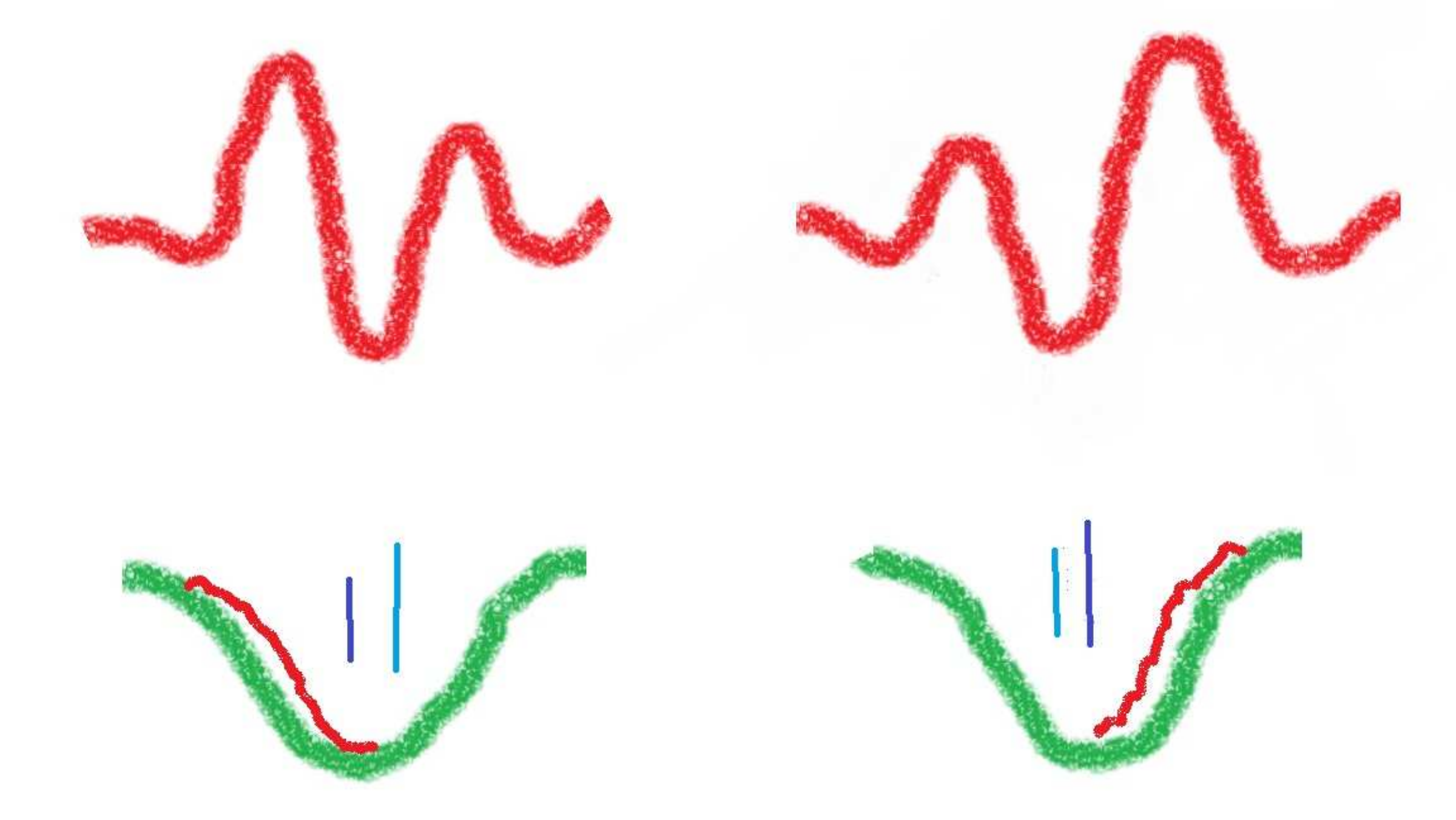}
\vspace{-2mm}
\caption{An illustration showing how the phase-locked $V/R$ changes
in a binary system can increase the true amplitude of the orbital RV changes
of seemingly absorption lines, partly filled by a  weak emission.}
\label{vr}
          \end{figure}
           \begin{figure}[t]
\centering
\includegraphics[width=95mm,angle=0]{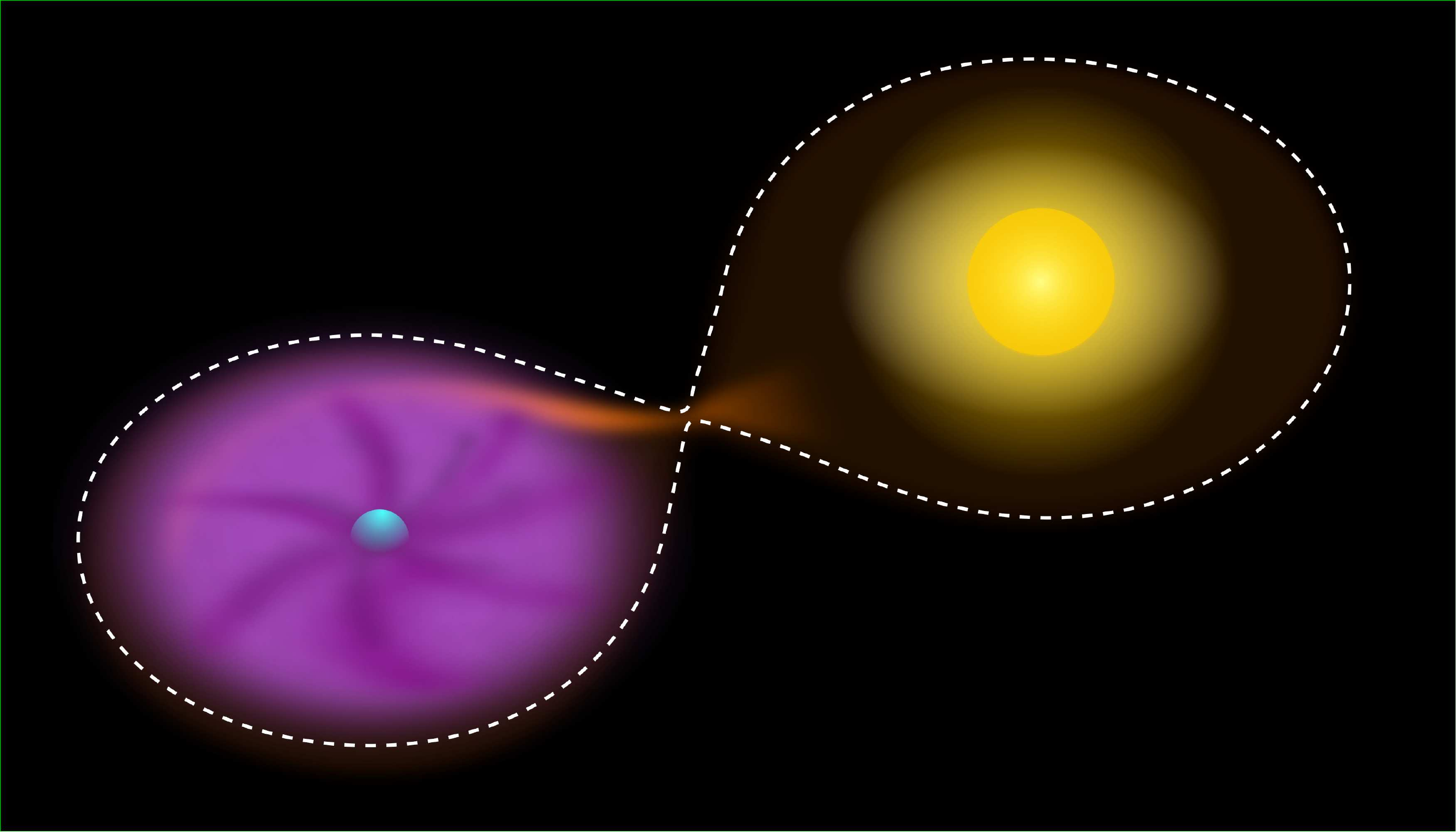}
\vspace{-2mm}
\caption{A cartoon of the system in our interpretation, showing the
F supergiant with an \ha emission-producing envelope, a cool disk
around the B-type secondary with some corotating structures, and a gas
stream
flowing from the primary towards the disk. (Visualisation by David Ond\v
rich.)}
\label{cartoon}
\end{figure}

One can consider two possible explanations:
\begin{itemize}
\itemsep -2pt
\item Either the wings of the \ha emission describe the orbital
motion of the F star correctly and the higher RV amplitude of absorption
lines
is due to their slight filling by emission combined with the phase-locked
$V/R$ changes as illustrated in Figure~\ref{vr} \citep[see also][]{bbin2003};
we remind that the $V$ peak of the emission is indeed statistically
stronger prior to eclipse, when also orbital RV is more positive; or
\item the optical centre of the \ha emission is closer to
the centre of gravity of the binary and the absorption-line RV amplitude
describes the orbital motion of the F star better than that of the
\ha emission.
\end{itemize}


\noindent An idea which comes to mind is:
What if the F primary is in the process of
{\sl episodic atmospheric mass transfer} each periastron passage via
the Roche-lobe overflow? This is a situation known for some Be+X-ray
binaries having eccentric orbits. Note that for \epe,
the periastron passage occurs close to
the primary mid-eclipse. Only the optically thin gas would be affected,
the photosphere of the F star being safely detached.
Appropriate modelling of such mass transfer should show
whether this would be able to maintain the dense disk around the secondary.

\vspace{-2mm}
\section*{Acknowledgements}
\vspace{-1mm}
We thank Drs. P.D.~Bennett and S.~Yang for the permission to use their DAO
spectra (analyzed previously in a joint study). We are very obliged
to Mr. David Ond\v rich for the cartoon of the system.
The research of PH, MW and PZ was supported
by the grant GA~\v{C}R P209/10/0715.



\end{document}